\documentclass[pre,twocolumn,showpacs,amsmath,amssymb]{revtex4}
\usepackage{graphicx}
\usepackage{dcolumn}
\usepackage{bm}

\begin{document}
\title{Effects of kink and flexible hinge defects on mechanical responses of short double stranded DNA molecules}
\author{Hu Chen}
\email[Email: ]{phych@nus.edu.sg}
\author{Jie Yan}
\email[Email: ]{phyyj@nus.edu.sg}
\affiliation{National University of Singapore, Department of Physics, 2 Science Drive 3, Singapore 117542}

\date{\today}

\begin{abstract}
We predict various detectable mechanical responses to the presence of local DNA defects which are defined as short DNA segments exhibiting mechanical properties obviously different from the 50 nm persistence length based semiflexible polymer model. The defects discussed are kinks and flexible hinges either permanently fixed on DNA or thermally excited. Their effects on extension shift, the effective persistence length, the end-to-end distance distribution, and the cyclization probability are computed using a transfer-matrix method. Our predictions will be useful in future experimental designs to study DNA nicks or mismatch base pairs, mechanics of specific DNA sequences, and specific DNA-protein interaction using magnetic tweezer, fluorescence resonance energy transfer or plasmon resonance technique, and the traditional biochemistry cyclization probability measurements.
\end{abstract}
\maketitle

\section{Introduction}

Single-molecule stretching experiments for micron size DNA and cyclization experiments for DNA larger than $230$ base pairs (bp) have shown that DNA can be considered as a homogeneous semiflexible polymer with an averaged bending persistence length $\sim 50$ nm \cite{Smith1992,Bustamante1994,Marko1995,Shore1983}. The bending energy is a quadratic function of local curvature in semiflexible polymer. On the other hand, the rigidity and local curvature of short DNA sequences are sensitively sequence-dependent \cite{Olson1998,Scipioni2002}. Understanding the micromechanics of such sequences is important since they regulate protein binding and DNA compaction \cite{Garcia2007a}. In this paper, a local DNA segment is defined as a mechanical defect if its elastic property differs from the 50 nm persistence length based semiflexible polymer model. A DNA defect can arise from sequence-dependent mechanical inhomogeneity, binding of a protein, or changes in DNA secondary structure.  A defect can be permanently  fixed at a specific location, or subject to thermal fluctuation. Hereafter we call the former a fixed defect, and the latter an excited defect. In the research, we study the how the defect changes several mechanical behaviors of short DNA. The changes may be detectable by experiments, and they reflect important information of the micromechanical properties of the defects.

The defects can also be introduced by protein binding which usually creates a local bending, or modifies the local bending rigidity, or both \cite{Geanacopoulos2001,Ali2001,Amit2003,Noort2004}. Defects created this way are not fixed but dynamic, subject to thermal fluctuation of protein binding or unbinding. In equilibrium, the fluctuation is determined by the interaction energy between the protein and DNA. In many cases, proteins prefer binding to some specific sequences with very strong interaction. In such cases the protein-binding induced defects can also be considered fixed.

Furthermore, defects may also arise from the changes in the secondary structures of DNA. For examples, a DNA nick or a mismatch bubble that contains one or a few mismatched base pairs may lead to a drastic change in the local curvature and rigidity of DNA \cite{Du2005a, Archer2006}. A DNA bubble that contains one or a few melted base pairs may also appear by thermal fluctuation. In fact, the defects arising from the changes in the secondary structure of DNA have recently been argued to affect the micromechanics of sharply bent DNA greatly \cite{Cloutier2004, Wiggins2006Nature, Shroff2005, Du2007}, and it has been argued that, in order to correctly describe the micromechanics of highly bent DNA conformations, the traditional 50 nm persistence length based semiflexible polymer model of DNA should be revised to include the excitation of such defects \cite{Yan2004, Wiggins2005, Du2007}. 

We will focus our interest in two types of defects: one enhances the local flexibility, and the other creates a local rigid bend. Hereafter we call the former a ``hinge'', and the latter a ``kink''. Traditional biochemistry methods to study such DNA defects are mainly based on gel electrophoresis for short DNA, in which the defect-resulted gel shift is observed and analyzed \cite{Kahn1994}, or based on the ``J-factor'' measurement in which the cyclization probability of short DNA is measured \cite{Kahn1992}. Recently, the rapid development of single-molecule manipulation and imaging methods has made it possible to study the defects more directly. This is because the defects can introduce detectable changes in DNA force response, DNA shapes, DNA cyclization probability, and DNA end-to-end distance distributions. In this paper, we predict the effects of the hinge and kink defects on these mechanical signals. 

Effects of kinks on force-extension curve of DNA have been discussed previously for a long DNA. It has been shown that there must be sufficient number of copies of the defects on the DNA in order to see a detectable shift \cite{Yan2003}. It was estimated that in order to see the effects, the line density of the defects on DNA should be at least $1$ per $100$ nm. As a result, force-extension curve is not suitable to study excited defects when the excitation energy is too large. For the same reason, to study the effect of a single fixed defect, the length of DNA has to be not longer than $\sim 100$ nm. Previous single-DNA stretching experiments were usually done for DNA of a few micro meters in length. For the reasons discussed above, although force-extension curve measurement is a powerful method to study non-specific interaction DNA-protein interactions, it is not suitable to study specific DNA-protein interaction when the specific protein binding sites are rare in DNA. In general, stretching a long DNA cannot detect the effects of defects that are rarely distributed. 

In order to study specific DNA-protein interaction or the mechanical property of single defects using force-extension curve measurement, one has to work with short DNA and compare with theoretical prediction at short length scales. Although such experiments were not reported previously, they are doable in principle. A possible design of such experiments is discussed in section IV. Theoretical predictions of force-extension curve were not reported at short lengths, perhaps due to the lack of experiments of stretching short DNA. Due to the important applications of stretching short DNA and the lack of corresponding theoretical predictions, we predict the force-extension curve of a 100 nm DNA, and the effects  of a single fixed kink or a flexible hinge in section III.A. We show that the shift in extension induced by a single defect is sufficiently large to be detected using single-molecule stretching instrument. 

The presence of defects also affects the end-to-end distance distribution of short DNA. Experimentally, several methods can accurately record the dynamical fluctuation of the end-to-end distance of DNA. The well-known fluorescence resonance energy transfer (FRET) technique is able to measure dynamical fluctuation of end-to-end distance of distances shorter than 10 nm, and it has been used to study the property of DNA bubble containing one or a few mismatched base pairs \cite{Archer2006}. Another very useful method is the recently developed plasmon resonance ruler that is able to measure dynamical fluctuation of distances ranged between $1-100 $ nm \cite{Sonnichsen2005}. This method is perhaps more powerful since it effects in a much wider range and it does not have the photo bleach problem. In order to make predictions for possible experiments using plasmon resonance ruler, we study the effects of defects on end-to-end distance distribution for a short DNA in section III.B. The study covers both fixed and excited hinge and kink, which is more complete than previously reported results \cite{Yan2004, Yan2005, Ranjith2005}.

As a special case of DNA end-to-end distance distribution, DNA cyclization probability can be measured using biochemistry methods. Experimentally, it has been used to study the bending and twisting rigidities of a single stranded DNA (ssDNA) gap of several base pairs \cite{Du2005a}. Theoretically, the looping probability of short DNA has been shown sensitive to the presence of defects. The effects of flexible hinge excitation and fixed defects on looping probability of short DNA have been studied previously \cite{Yan2004, Wiggins2005, Du2005a, Ranjith2005}. In section III.(C-D), in addition to re-confirming the previous predictions, we compute the effects of kink excitation, the multiple defects, and relative positions of the defects which have not been discussed previously. Our results show that excited kink has similar effect on J-factor as excited hinge previously studied \cite{Yan2004}, suggesting an alternative mechanism to explain the unexpected j-factor of $\sim$ 100 bp DNA reported in \cite{Cloutier2004}. This excited kink mechanism is different from the kinkable elastic polymer mechanism previously proposed \cite{Wiggins2005}, where the excited defect is a hinge of zero rigidity. 

In section III.D, we also predict the end meeting angle dependence of DNA looping and its interaction with the presence of defects. Such predictions are important in studies of DNA loops where the ends are bridged by a protein, since such DNA looping proteins may likely impose a specific requirement on the meeting angle of DNA. Previously, it has been demonstrated that the protein size greatly affects the looping probability \cite{Douarche2005}, but the effects of meeting angle has not been studied. Our results show that the looping probability of a 40 nm DNA is extremely sensitive to the meeting angle and to the presence of defects. 

The computation of the force-extension curve, the end-to-end distance distribution, and the cyclization probability of DNA are done using the transfer-matrix methods previously developed in \cite{Yan2004, Yan2005}. We believe our predictions will be useful for future experiments designed to study the micromechanical properties of the defects. 

\section{Theory and Model based on Transfer Matrix Method}\label{theory}


We consider the discrete polymer model consisting of $N$ segments with equal length $b$. In such a model, the conformation of DNA is described by the orientations of the segments, $\hat{\bf t}_i$, where $i=1,\cdots, N$, and the conformational energy is carried by the bending of the vertices connecting adjacent segments. The energy in unit of $k_B T$ of this model is a summation of the vertex energies:
\begin{equation}
E = \sum\limits_{i=1}^{N-1} E_i(\hat{\bf t}_{i},\hat{\bf t}_{i+1}).
\end{equation}

For semiflexible polymer model of DNA, $E_i= {a \over 2}\left( \hat{\bf t}_{i+1}-\hat{\bf t}_i \right)^2$, where $a=A/b$ is a dimensionless quantity describing the rigidity of the vertex. The parameter $A$ is the persistence length of the semiflexible polymer which describes the local rigidity of the polymer. In order to approximate the continuum semiflexible model, it is required to choose a segment length $b\ll A$. For DNA, it has been accepted that $A=50$ nm under normal experimental conditions \cite{Smith1992,Bustamante1994,Shore1983}. Our computations described in the following sections are based on a choice of $b=1$ nm, thus $a=50$ for DNA in the 50 nm persistence length based semiflexible polymer model. 

The site-dependent vertex energy $E_i(\hat{\bf t}_{i},\hat{\bf t}_{i+1})$ is able to describe fixed defects. Two kinds of mechanical defects are discussed in the paper: 1) the flexible hinge defect $E_{i,hinge}(\hat{\bf t}_{i+1},\hat{\bf t}_i) = {a'\over 2}\left( \hat{\bf t}_{i+1}-\hat{\bf t}_i \right)^2$, where $a'$ is much smaller than the vertex rigidity of B-form DNA, and 2) the kink defect $E_{i,kink} = {a\over 2}(\hat{\bf t}_i \cdot \hat{\bf t}_{i+1}-\gamma)^2$, where $\gamma=\cos(\theta)$ and $\theta$ is the preferred (lowest-energy) kink angle. These defect energy forms have been used in some of our previous studies \cite{Yan2003, Yan2004, Yan2005}. In the computation, we choose $a'=1$ for a flexible hinge vertex, and ($\theta = \pi/2$, a=50) for a kink vertex. 

To describe the excited state, a vertex can be made to have two states, indicated by an index $n_i$: $n_i=0$ for B-form state, and $n_i=1$ for defect state. With the excitation, the vertex energy becomes:
\begin{equation}
\begin{array}{ll}
E_i(\hat{\bf t}_i,\hat{\bf t}_{i+1} ) &=
\delta_{n_i,0}E_{i,0}(\hat{\bf t}_i,\hat{\bf t}_{i+1} )\\
&+ \delta_{n_i,1} (E_{i,1}(\hat{\bf t}_i,\hat{\bf t}_{i+1}) + \mu),\
\end{array}
\end{equation}
where $E_{i,0}$ is the B-form DNA vertex energy, and $E_{i,1}$ is the defect energy. The parameter $\mu$ controls the probability that a defect forms, and represents the free
energy cost of generating a local defect\cite{Yan2004,Yan2005}.


A transfer-matrix method has been developed to compute the end-to-end distance distribution, the looping probability, and the force-extension curve \cite{Yan2005}. The partition function of conformations subject to certain constraints can be calculated by the inverse Fourier transformation of the partition function in wave number space $Z_{\bf k}$. $Z_{\bf k}$ can be computed by a direct order product of the site-dependent matrices, so the finite size effect of short DNA, sequence-dependence, and the boundary condition are automatically included. The probability of specific DNA conformation is the ratio of the constrained partition function and total partition function. To calculate the extension of DNA subject to certain stretching force, a force induced energy $\sum\limits_{i=1}^{N-1} {-bf \hat{\bf z} \cdot \hat{\bf t}_i }$ is added into the energy equation. Extension of DNA is obtained from the derivative of the free energy with respect to force. Please find the details of the transfer-matrix method in \cite{Yan2005}.

\section{DNA mechanical responses to the defects}

\subsection{DNA extension under tensile force}
In a single molecular manipulation experiment, a tensile force can be applied to DNA and the resulted extension of DNA can be measured. The mechanical properties of DNA can be derived from the force-extension curve \cite{Bustamante1994,Marko1995}. In such experiments, the DNA usually has a contour length of a few micrometers. In \cite{Yan2003}, it was shown that in order to see detectable effects of defects on the force-extension curve, the density of the defects on DNA must be sufficiently large. The minimal density of kink defects was shown to be around one per 100 nm \cite{Yan2003}. Because of this, one does not expect to see the effects of excitation of defects if the excitation energy is too high \cite{Yan2005}, or the effects of a single fixed defect if the DNA is too large. In this section, we predict the detectable effects of single fixed defects on the force-extension curve of a short DNA of $100$ nm.

\begin{figure}[htbp]
\centering
\includegraphics[width=8cm]{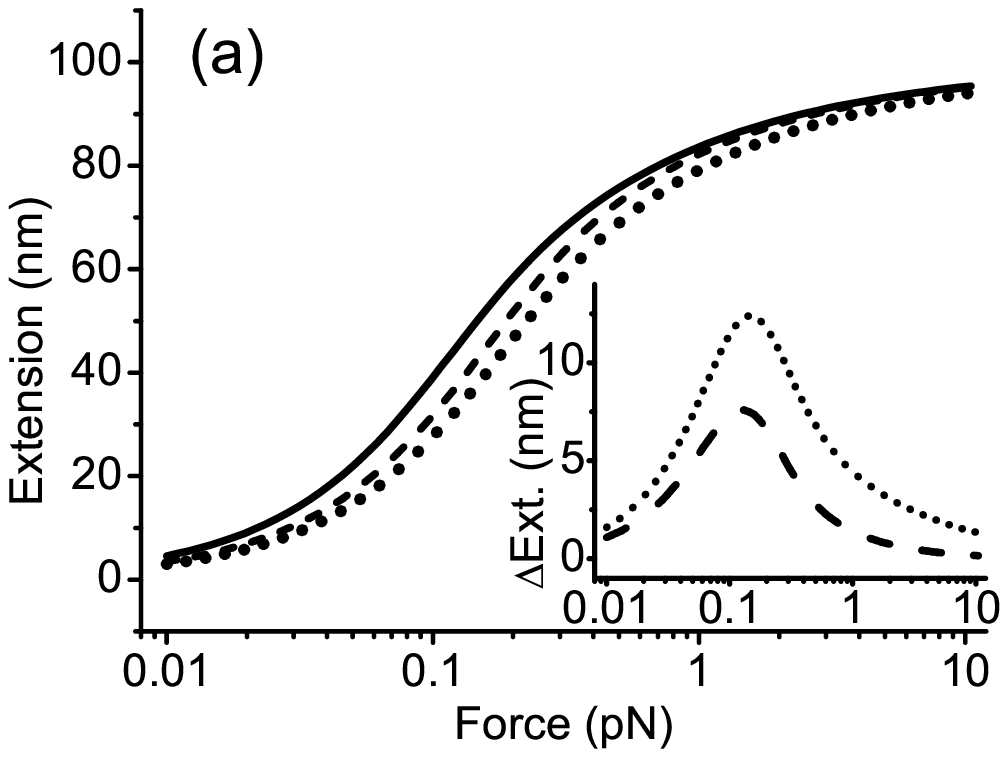}
\includegraphics[width=8cm]{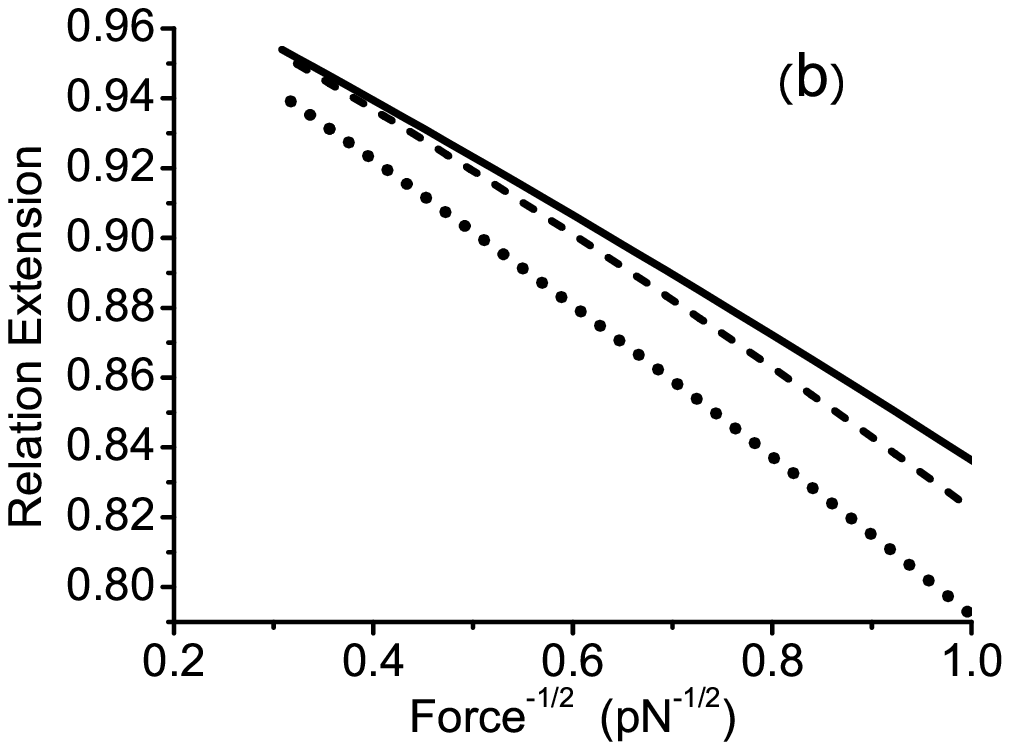}
\caption{The effects of fixed defects on force-extension curve measurements. (a) The force-extension curve of a 100 nm DNA that does not contain defects (solid curve), contains a single flexible hinge defect at center (dashed curve), and contains a single $90^o$ kink defect at center (dotted curve). Inset shows the magnitude of the shift of the defect-containing DNA curves from the defect-free DNA curve. (b) The slopes are used to compute effective persistence lengths, which are found to be $\sim 36$ nm for the defect-free DNA, $\sim 30$ nm for the fixed flexible hinge containing DNA, and $\sim 22$ nm for the fixed kink containing DNA.}
\label{fx}
\end{figure}

The force-extension curve may be computed using the site-dependent transfer-matrix method developed in \cite{Yan2005}. Fig. \ref{fx} shows the force-extension curve of a $100$ nm long DNA that contains a single defect in the middle. From the figure one can see a single fixed defect results in detectable shift of the force-extension curve. Inset of Fig. \ref{fx} (a) shows the shift of the defect-containing DNA curve from the defect-free DNA curve. The largest shift is $\sim 7$ nm for DNA containing a flexible hinge, and $\sim 12$ nm for DNA containing a kink, at a force $\sim 0.14$ pN. Such shifts are detectable in single molecule experiments. Alternatively, one can study the effects of the defects by measuring the effective persistence of DNA. The effective persistence length may be determined by fitting the well-known force-extension curve at large force approximation \cite{Marko1995}: ${<z>/L} = 1-\sqrt{k_B T/ 4Af}$, where ${<z>/L}$ is the average extension scaled by the contour length, $f$ is the tension along $\hat{z}$ direction, and $A$ is the effective persistence length. Fig. \ref{fx}(b) shows the results: effective persistence of the defect-free DNA determined by the slope fitting is $A\sim 36$ nm, apparently smaller than the expected $50$ nm. This reduced value is due to the finite-size effect which was recently discussed in \cite{Seol2007}. The slope fitting for defect-containing DNA gives $A\sim 30$ nm for DNA containing a single flexible hinge, and $A\sim 22$ nm for DNA containing a single $90^o$ kink. Longer DNA with contour length $500$ nm is also studied, where the effects of the defects become much less detectable. The effective persistence length is $A\sim 46$ nm for defect-free DNA, $A\sim 44$ nm for DNA containing a flexible hinge, and $A\sim 41$ nm for DNA containing a $90^o$ kink.

\begin{figure}[htbp]
\centering
\includegraphics[width=8cm]{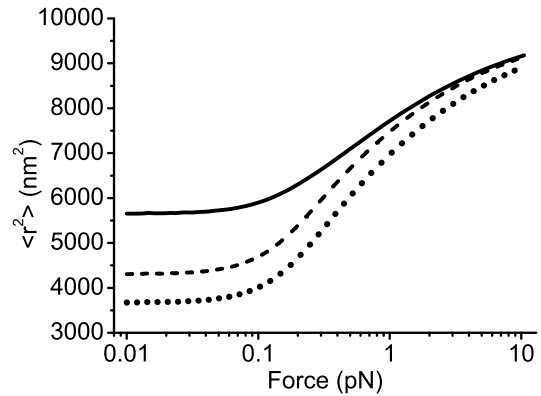}
\caption{$<r^2>$ as functions of stretching force for defect-free and defect-containing DNA. The symbols are the same as in Fig. \ref{fx}. }
\label{r2f}
\end{figure}

When tension approaches zero, the average extension of DNA also approaches zero. As such, at small tension, the radius of end-to-end distance ($r$) is a better measurable quantity \cite{Neumann2003}. Subject to a tension along $\hat{z}$ direction, $<r^2>$ is determined by the following equation:
\begin{equation}
<r^2> =<z>^2 + \sigma^2_z + \frac{2 k_B T <z>}{f},
\end{equation}
where $\sigma^2_z$ is the variance of DNA extension which can be obtained from the secondary derivative of free energy. The effects of defects on $<r^2>$ are shown in Fig. \ref{r2f}. At small force, $<r^2>$ does not vanish as expected, and it is determined by the radial end-to-end distance distribution of a tension-free DNA which will be discussed in the next section.

\subsection{End-to-End distance distribution}

An approximate analytical expression has been derived for the radial end-to-end distance distribution of short semiflexible polymers \cite{Wilhelm1996}. By using the transfer matrix method developed in \cite{Yan2005}, the distribution of the end-to-end distance vector (${\bf R}$) of a polymer subject to defect excitation or containing fixed defects can be calculated. In many cases the vector distribution $\rho({\bf R})$ does not depend on the direction of ${\bf R}$, and in such cases the scalar end-to-end distance distribution is simply $4 \pi R^2 \rho({\bf R})$.

\begin{figure}[htbp]
\centering
\includegraphics[width=8cm]{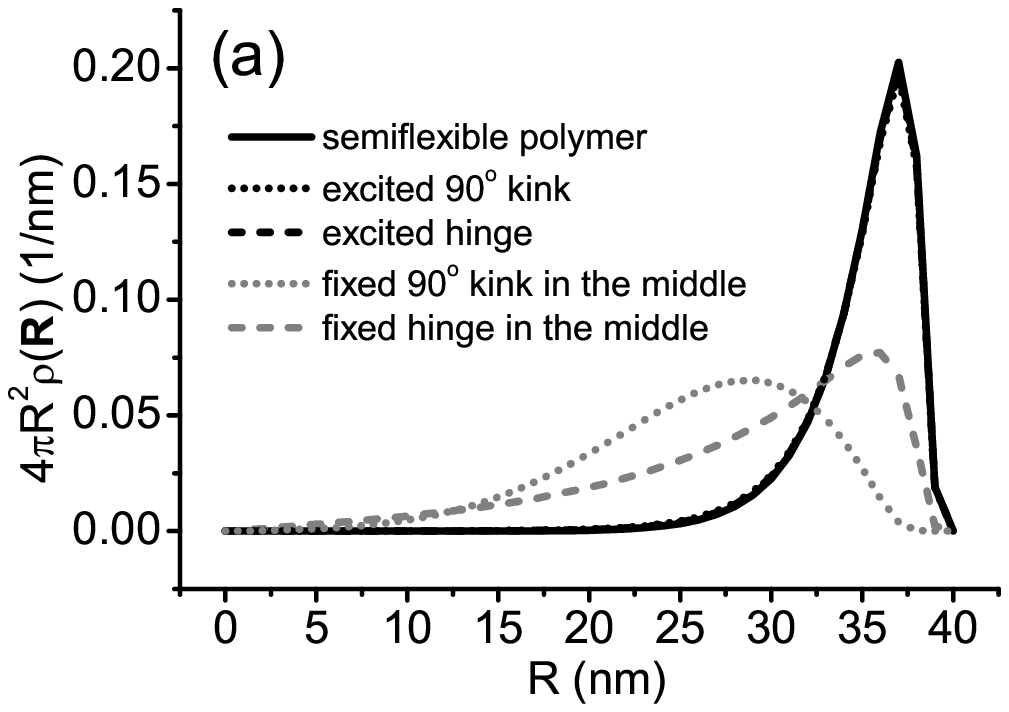}
\includegraphics[width=8cm]{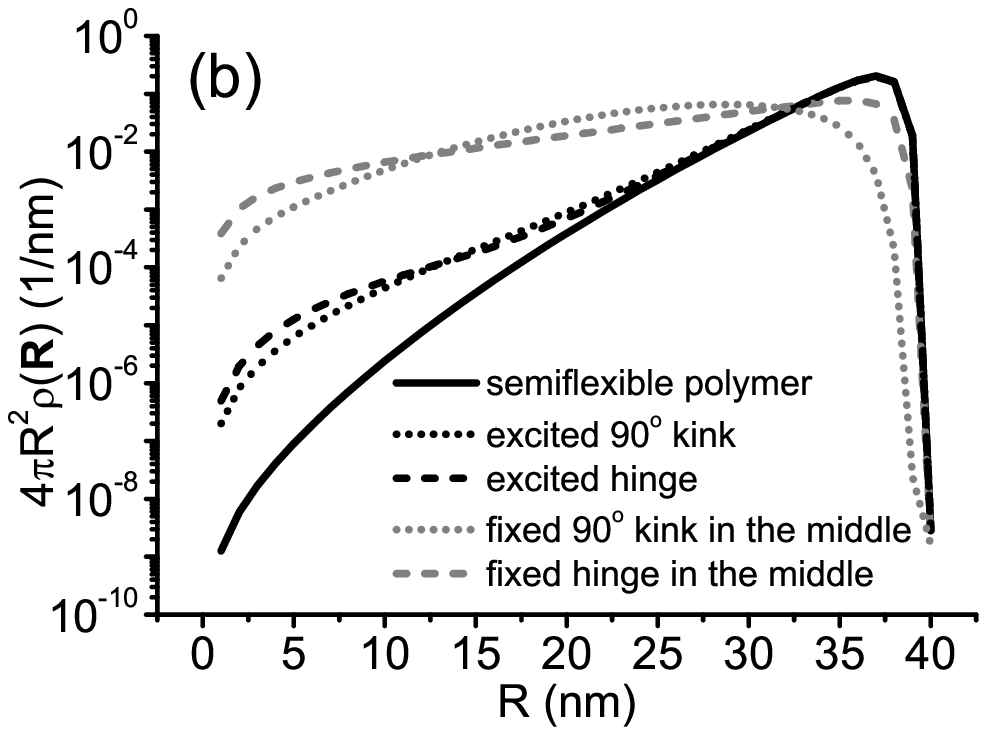}
\caption{Effects of defects on scalar end-to-end distance distribution of a $40$ nm long DNA plotted in linear scale (a) and logarithm scale (b). In (a), curves for excited defects are nearly overlapped with defect-free DNA. The fixed defects are located in the middle of the DNA. For excited defects, excitation energy for kink is 10 $k_B T$, and that for flexible hinge is 11 $k_B T$. }
\label{figR}
\end{figure}

Fig. \ref{figR} shows how fixed and excited defects affect the scalar end-to-end distance of a $40$ nm DNA. In the studies of the excitation of defects, the excitation energy for kink is 10 $k_B T$, and that for flexible hinge is 11 $k_B T$. Clearly, the distribution of end-to-end distance of short DNA is extremely sensitive to the fixed defects as shown in Fig. \ref{figR} (a). The peak of the distribution moves to smaller end-to-end distance, and becomes wider. However, the effects of the excitation of the defects are much less obvious in Fig. \ref{figR} (a). This is because the distribution is most sensitive to the defects only when $R$ is very small where the excited defects begin to dominate the bending. To see it, we replot the distribution using logarithm scale as shown in Fig. \ref{figR} (b). Our results show that, when $R<10$ nm, $4 \pi R^2 \rho({\bf R})$ for the DNA subject to excitations of defects is about two orders of magnitude larger than that of a defect-free DNA.

The position of the fixed defect affects the end-to-end distance distribution.
We computed the positional effect of a fixed $90^o$ kink and a fixed flexible hinge for a $100$ nm long DNA. For both defects, their effects are maximized when they are in the middle of DNA. When the defect moves from one end to the center of the DNA, the peak of the distribution moves to shorter end-to-end distance, and the distribution becomes wider. This observation is in agreement with the computation done in \cite{Ranjith2005} where the positional dependence of a fixed flexible hinge was studied. 

\begin{figure}[htbp]
\centering
\includegraphics[width=8cm]{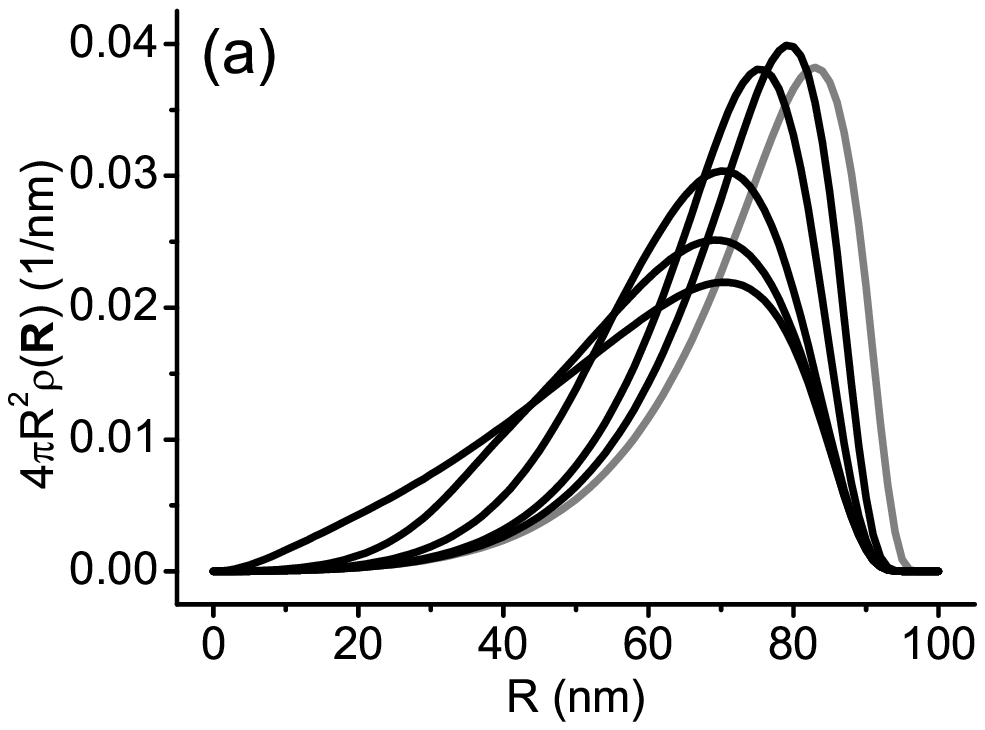}
\includegraphics[width=8cm]{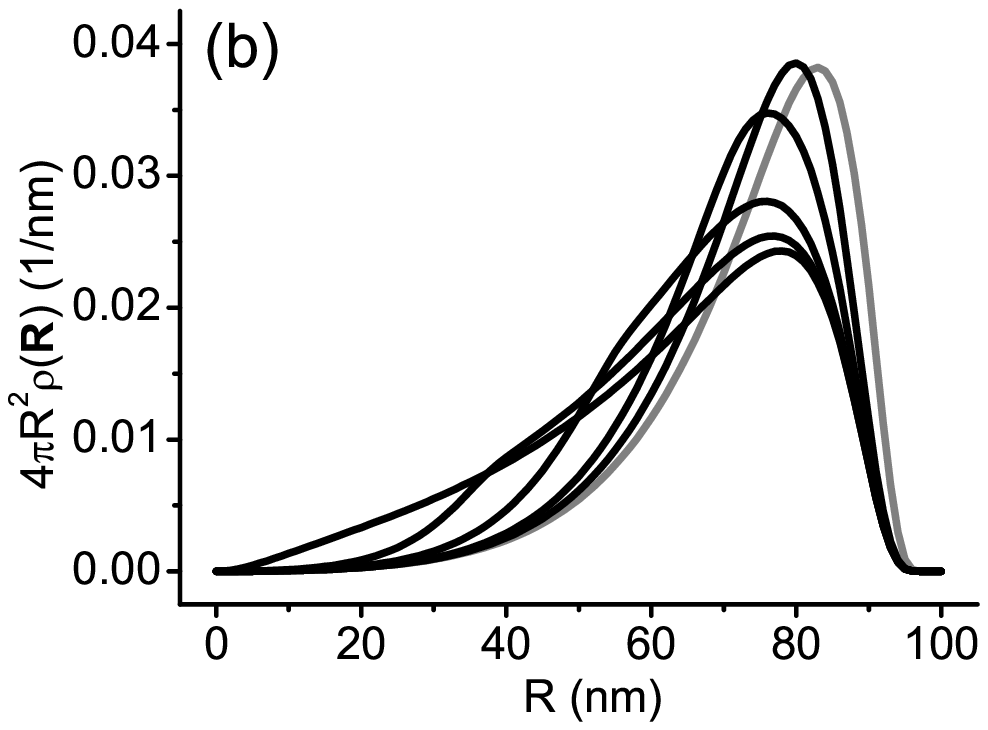}
\caption{The effects of the positions of fixed defects on the scalar distribution of end-to-end distance of a $100$ nm long DNA. (a) The positional effect of a fixed $90^o$ kink. (b) The positional effect of a flexible hinge. In (a) and (b), the reference distributions of the defect-free DNA are displayed in gray color. The dark curves, from top to bottom, are results computed for defect locations at 5, 10, 20, 30, 50 nm from one end of DNA, respectively. }
\label{figRpos}
\end{figure}

\subsection{J-factor measurement of DNA molecules}

The previous section shows the effects of defects on the scalar end-to-end distance distribution $4 \pi R^2 \rho({\bf R})$. Their effects are found much enhanced when $R$ is small, and one may expect the effects are maximized at $R=0$ where DNA loop forms. Experiments can be carried out in such a way that the rate of cyclization of DNA is proportional to $\rho_{//}({\bf 0})$ which is just $\rho({\bf 0})$ under a constraint that the two DNA ends are parallel to each other \cite{Shore1981}. 
The quantity measured in such experiments is the ``J-factor'', which is related to $\rho_{//}({\bf 0})$ by $J = {4\pi \over N_A}\rho_{//}({\bf 0})$, where $N_A$ is Avogadro's number. This quantity, usually expressed in units of mol/liter (M), is a measure of the equilibrium concentration of one end of the DNA at the position of other end. Several methods have been developed to compute the J-factor for semiflexible polymers \cite{Yamakawa1984, Vologodskii2000, Zhang2003} 

In the transfer-matrix computation, due to the parallel boundary condition, the constrained partition function of circular conformations is the trace of the ordered product of the site-dependent matrices \cite{Yan2005}. Because of the cyclic property of trace, the cyclization probability of DNA containing a single fixed defect does not depend on the location of the defect. The details of the transfer-matrix computation of DNA cyclization probability is described in \cite{Yan2005}. The effects of excitation of flexible hinge and a single fixed kink on J-factor are discussed there. In this section we report a more complete computation for the effects of the defects, including the kink excitation and multiple defects which have not been discussed earlier. 

\begin{figure}[htbp]
\centering
\includegraphics[width=8cm]{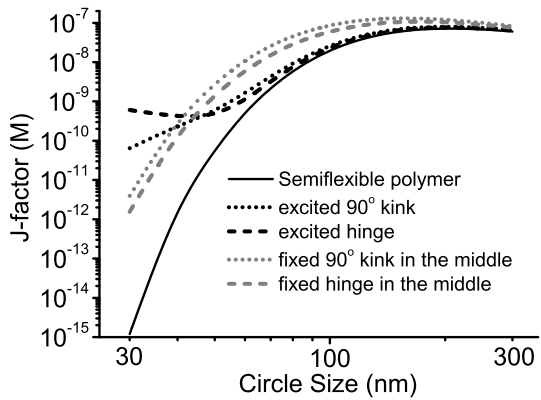}
\caption{J-factors as functions of DNA length. The fixed defects are located in the middle of DNA. For excited defects, the excitation energy for kink is 10 $k_B T$, and that for flexible hinge is 11 $k_B T$. } 
\label{figJ}
\end{figure}

Fig. \ref{figJ} shows the effects of fixed or excited defects described previously on DNA J-factor. For long DNA, the cyclization of DNA is dominated entropically, so the rarely distributed defects do not affect the J-factor much. On the other hand when DNA is shorter than $\sim 100$ nm, DNA bending energy begins to dominate the cyclization of DNA, and the effects of defects becomes enhanced since they greatly reduce the free energy of DNA circles. From Fig. \ref{figJ}, the effect of a fixed defect becomes noticeable when DNA is less than around $200$ nm, and becomes greater when DNA is shorter. Comparing with a single fixed defect, the effect of an excited defect becomes noticeable at a shorter length ($\leq 80$ nm). This is because there is a penalty of excitation energy of 10 $k_BT$ for kink or 11 $k_B T$ for hinge to have an excited defect. When DNA is shorter than $\sim 40$ nm, DNA with excited defects even begin to have J-factor larger than DNA with a single fixed defect, indicating that more than one defect is excited to further relax sharp bending of the mini circular DNA. Interestingly, J-factor of DNA with excited flexible hinge defect increases with decreasing DNA contour length when it is shorter than 40 nm. This phenomena can be easily understood: when there are two flexible hinges already excited at optimal positions to allow a direct folding to form DNA loop satisfying the parallel boundary condition, the looping is dominated by how easy the two nearly ``freely-joint'' segments can find each other. Obviously, the smaller the circle size, the larger the chance the two ends meet. As shown in \cite{Yan2004,Wiggins2005}, excited flexible hinge can give high J-factor for short DNA circles. Fig. \ref{figJ} shows that excited rigid $90^o$ kink also has similar effect. 

\begin{figure}[htbp]
\centering
\includegraphics[width=8cm]{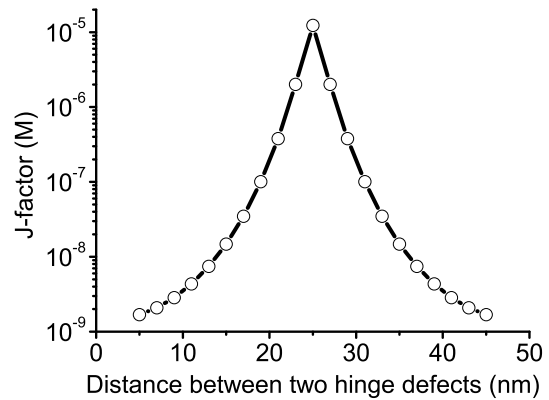}
\caption{The effect of separation of two fixed flexible hinge defects on J-factor of a $50 $ nm DNA.}
\label{J2bub}
\end{figure}

When there are more than one fixed defects along DNA, J-factor of such a polymer then depends on the locations of the two defects. Fig. \ref{J2bub} shows the how J-factor depends on the separation between two fixed flexible hinge defects on a $50 $ nm DNA. Clearly, J-factor is very sensitive to the relative position of the two defects. It increases about 4 orders of magnitude when the separation between two defects increases from near zero to half of the circle size, $25 $ nm. The optimal separation is half of the circle size, because this separation allows a direct folding for loop formation while satisfying the parallel boundary condition. The looping probability depends only on the separation between the two defects in the loop conformation, which is also a result of the cyclic property of trace. 

\subsection{Looping probability under free boundary condition and dependence on meeting angle}

The J-factor measurement discussed in the previous section essentially measures the cyclization probability with a parallel boundary condition. DNA loops can also form under other boundary conditions. For example, a protein mediated loop may impose a specific meeting angle constraint on the two DNA ends. If the interaction of the ends does not impose a strong orientational requirement, then we call it ``free boundary condition''. The effects of excitation of flexible hinge and a single middle-located fixed kink under the free boundary condition were also discussed in \cite{Yan2005}. The effect of the position of a single fixed flexible hinge on looping probability of a short DNA under free boundary condition was computed in \cite{Ranjith2005}. Here we report a more complete computation including the effects of fixed flexible hinge on a larger length range, the excitation of kinks, the positional dependence of kink, and the dependence on meeting angle constraint which have not been discussed earlier. 

\begin{figure}[htbp]
\centering
\includegraphics[width=8cm]{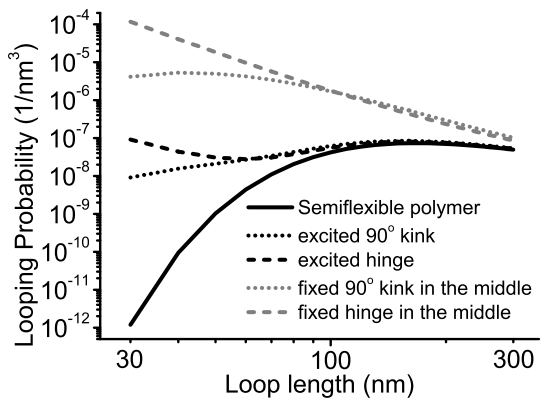}
\caption{Effects of defects on DNA looping under free boundary condition. The fixed defects are in the middle of the DNA. For excitation of defects, the excitation energy for kink is 10 $k_B T$ and that for flexible hinge is 11 $k_B T$. }
\label{looping}
\end{figure}

As shown in Fig. \ref{looping}, an interesting observation is that the looping probability of DNA with a middle-located fixed flexible hinge scales with loop length with a power law. This is because the hinge basically breaks the short DNA into two nearly freely joint segments. If the hinge is completely free to rotate and the two identical arms are infinite rigid, the looping probability scales inversely with the volume which the ends can sample. As a result, one expects $P_{looping}\propto 1/L^{3}$. In our computation, the flexible hinge has a finite bending rigidity and the two arms are not infinite rigid, leading to a slightly different scaling: $P_{looping}\propto 1/L^{3.5}$. The curve corresponding to the excitation of flexible hinges displays similar power law when DNA length is shorter than 40 nm. This is because when DNA becomes very short, a flexible hinge is always excited in the middle of the DNA to allow a direct folding to meet the two ends. The curve of the fixed kink does not display a simple power law. The large rigidity of the kink ($a=50$) leads to decreasing looping probability when the length is getting shorter than $\sim 40 $ nm.

\begin{figure}[htbp]
\centering
\includegraphics[width=8cm]{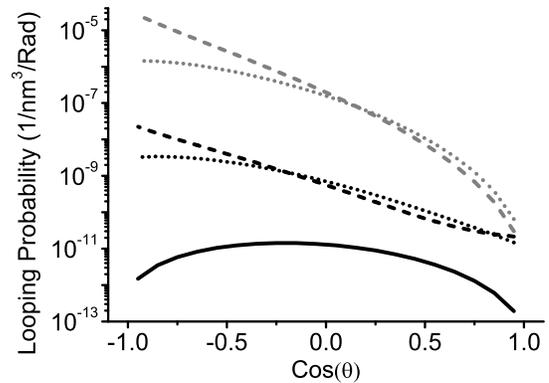} 
\caption{The dependence of looping probability on the meeting angle $\theta$ between the first and last tangent vector for 40 nm DNA. $Cos(\theta) =1$ for parallel boundary condition, while $Cos(\theta) =-1$ for anti-parallel boundary condition. The symbols are the same as in Fig. \ref{looping}. }
\label{meeting}
\end{figure}

Fig. \ref{meeting} shows the looping probabilities as functions of the meeting angle for defect-free DNA, excited and fixed defect-containing DNA with a contour length of $40 $ nm. For a perfect semiflexible polymer, the looped conformation with lowest bending energy is like a teardrop with meeting angle around 1.8 $rad$ \cite{Sankararaman2005}. This is confirmed by our calculation: the looping probability of defect-free DNA is maximal when meeting angle is around 1.8 $rad$. The defects increase the looping probability under anti-parallel boundary condition ($Cos(\theta) =-1$) the most while that under parallel boundary condition ($Cos(\theta) =1$) nearly the least. This is because a single defect permanently located or excited at the middle is sufficient to allow DNA loop formation under anti-parallel condition, while under the parallel condition, it requires up to $2$ defects to allow DNA to be looped. As such, under free boundary condition, defected DNA tends to form loops with anti-parallel meeting angle. In the presence of non-specific DNA crosslinking factors, such defect formation favors the formation of hairpin structures. 

\begin{figure}[htbp]
\centering
\includegraphics[width=8cm]{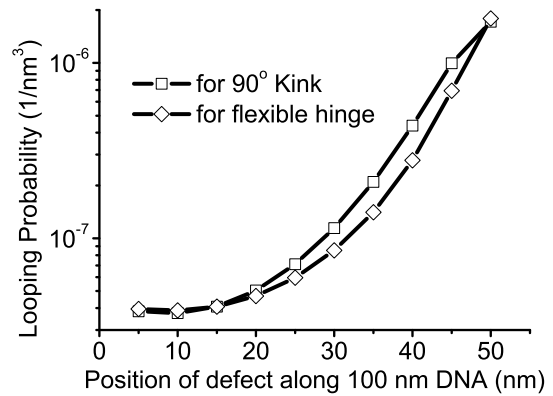}
\caption{Dependence of looping probability of a $100$ nm DNA on the positions of single fixed defects.}
\label{loop_p}
\end{figure}

Unlike the parallel boundary condition, looping under free boundary condition depends on the location of a single fixed defect. This is because the partition function is not the trace of the product of the matrices under the free boundary condition, as shown in \cite{Yan2005}. Fig. \ref{loop_p} shows the effects of the locations of single kink and single flexible hinge defects for looping of a $100$ nm DNA. In agreement with the observation in \cite{Ranjith2005}, the optimal defect position is the center of DNA. 

\section{Discussion and conclusion}

We have computed the effects of kink and flexible hinge defects on various mechanical responses of short DNA. We showed that the defects can lead to detectable shift in extension of a $\sim 100$ nm long DNA subject to a constant tension using a magnetic tweezer. We also showed that the effects of fixed defects can be detected by the changes in effective persistence lengths. To obtain the persistence length, one has to obtain accurate measurement for the tension and the extension. In magnetic tweezer, the tension is determined by the transverse fluctuation of one end of DNA and the extension of DNA. For short DNA, the relative errors in the measurements of both quantities usually become larger, leading to some potential technical difficulties to measure the effective persistence length with high accuracy. One possible method to overcome this difficulty is to use a long DNA handle which tethered between a paramagnetic particle and a non-paramagnetic particle. The non-paramagnetic particle is attached to the end of the short DNA being studied. The force is now applied to the paramagnetic particle, and it can be accurately determined using the long DNA handle. The extension of the short DNA is determined by the position of the non-paramagnetic particle. Using such a design, the accurate force-extension curve and the effective persistence length of a short DNA may be determined. 

We then computed the effects of defects on the end-to-end distance distribution of a 40 nm DNA. Such measurements can be done by recording the dynamical plasmon resonance wavelength of two 40 nm golden particles attached to the ends of DNA \cite{Sonnichsen2005}. This technique was reported able to monitor the dynamical changes in DNA end-to-end distance over a range of 1-100 nm. 
As such, our predictions of the end-to-end distance distribution of the 40 nm DNA are experimental detectable. 
As shown in Fig. \ref{figR}, both the fixed and excited defects affect the distribution, particularly when the end-to-end distance is shorter than 10 nm. If the DNA being studied is shorter than 10 nm, then the end-to-end distance fluctuation can also be measured by FRET technique. The dynamical end-to-end distance fluctuation measured by plasmon resonance or FRET can be used to study the dynamical binding of protein to DNA, the mechanical property of specific DNA sequences (for example, the nucleosome positioning sequences), the mechanical property of highly bent DNA, and other DNA defects such as DNA nicks and DNA mismatches.

Although there have been many theoretical computations for effects of defects on DNA cyclization and looping probability, we still include a section discussing it in order to make the research more complete. In addition to confirming a few previously reported predictions, some of our computations have not been discussed before. These include the effect of a fixed flexible hinge over a larger length range, the effect of multiple defects and excitation of rigid kinks, as well as the dependence on meeting angle. In the computation of the effect of a fixed defect on DNA J-factor measurement, one can see in Fig. \ref{figJ} that a single fixed flexible hinge takes a detectable effect up to loop size of $\sim 200$ nm ($600$ bp). We note that the effect of a single fixed flexible hinge on J-factor measurement is equivalent to the effect of a flexible meeting angle boundary condition that imposes an energy of the meeting angle ($\theta$) of the ends, $E(\theta)={a\over 2} \theta^2$, on J-factor measurement of a defect-free DNA of the same length. Thus, one would predict that the effect of such an elastic meeting angle constraint can affect the J-factor up to a looping size  $\sim 200$ nm if taking $a=1$. Recently, the flexible meeting angle boundary condition was proposed \cite{Tkachenko2007} to explain the discrepancy between the J-factor measurements of $\sim 100$ bp DNA reported in \cite{Cloutier2004} and \cite{Du2005}. In this mechanism, the possible effect of the two nicks after the annealing of two DNA overhangs and before the ligation reaction is considered. The nicked, annealed overhangs of 4 bp were assumed to be a flexible hinge in the mechanism. By varying the rigidity of the defect, it is shown that this mechanism is able to explain the both experiments using $a=1$ for J-factor reported in \cite{Cloutier2004} and $a=20$ for J-factor reported in \cite{Cloutier2004}. Noticing the equivalence between this mechanism and the fixed flexible hinge in their effects on J-factor, therefore, this mechanism does not explain J-factor measurements for DNA larger than $230$ bp, where all the measurements are in good agreement with predictions by 50 nm persistence length based semiflexible model \cite{Shore1983, Cloutier2004, Du2005}. 

The effects of meeting angle constraint on DNA looping have not been discussed previously. In Fig. \ref{meeting}, we show that the looping probability of a 40 nm defect-free DNA is sensitive to meeting angle, with a variation of about two orders of magnitude depending on the value of meeting angle. The sensitivity of DNA looping to meeting angle has important biological implications, since it is required to understand the looping efficiency of DNA with the ends bridged by proteins that impose a specific constraint on the meeting angle. As mentioned in introduction, another important factor affecting DNA looping efficiency is the size of the protein bridge \cite{Douarche2005}. We also show that, in the presence of defects, the looping probability in the whole angle range can be shifted up by at least two orders of magnitude at $\theta=0$, and up to seven orders of magnitude at $\theta=\pi$. This suggests that DNA with a defect tends to form loops in anti-parallel configurations. It implies formation of hairpin structure when the loop is stabilized by DNA bridging proteins such as HN-S. A defect would initiate formation of a seeding loop with a near anti-parallel boundary condition which is stabilized by a single bridging protein, followed by progressing into a hairpin structure by adding more proteins sequentially.  

All our computations described in the paper are done based on the transfer-matrix method developed previously \cite{Yan2005}. This method allows us to study the sequence-dependent properties, fixed or excited defects, and different boundary conditions. There is, however, a drawback in the current transfer-matrix method: there is no easy way to include the twist constraint into this method, since with twist we no longer have localized interaction which only depends on adjacent orientations of DNA. Since twist constraint affects DNA looping probability and shapes of DNA loops greatly for small DNA, one of our future researches is to find a way to include the twist constraint into the transfer-matrix computation.

Finally, we believe our predictions will find its applications in designs of future experiments to study DNA defects, specific DNA sequences, and specific DNA-protein interaction, using magnetic tweezer, FRET or plasmon resonance techniques, and the traditional biochemistry cyclization probability measurement.

\section*{Acknowledgment}
This research was supported by the Ministry of Education of
Singapore through Grants No. R144000143112 and No. R144000171712.
The authors thank Prof. John Marko, Rob Phillips, Philip Nelson, and Richard Neumann for valuable discussions.


\begin{thebibliography}{36}
\expandafter\ifx\csname natexlab\endcsname\relax\def\natexlab#1{#1}\fi
\expandafter\ifx\csname bibnamefont\endcsname\relax
  \def\bibnamefont#1{#1}\fi
\expandafter\ifx\csname bibfnamefont\endcsname\relax
  \def\bibfnamefont#1{#1}\fi
\expandafter\ifx\csname citenamefont\endcsname\relax
  \def\citenamefont#1{#1}\fi
\expandafter\ifx\csname url\endcsname\relax
  \def\url#1{\texttt{#1}}\fi
\expandafter\ifx\csname urlprefix\endcsname\relax\def\urlprefix{URL }\fi
\providecommand{\bibinfo}[2]{#2}
\providecommand{\eprint}[2][]{\url{#2}}

\bibitem[{\citenamefont{Smith et~al.}(1992)\citenamefont{Smith, Finzi, and
  Bustamante}}]{Smith1992}
\bibinfo{author}{\bibfnamefont{S.~B.} \bibnamefont{Smith}},
  \bibinfo{author}{\bibfnamefont{L.}~\bibnamefont{Finzi}}, \bibnamefont{and}
  \bibinfo{author}{\bibfnamefont{C.}~\bibnamefont{Bustamante}},
  \bibinfo{journal}{Science} \textbf{\bibinfo{volume}{258}},
  \bibinfo{pages}{1122} (\bibinfo{year}{1992}).

\bibitem[{\citenamefont{Bustamante et~al.}(1994)\citenamefont{Bustamante,
  Marko, Siggia, and Smith}}]{Bustamante1994}
\bibinfo{author}{\bibfnamefont{C.}~\bibnamefont{Bustamante}},
  \bibinfo{author}{\bibfnamefont{J.~F.} \bibnamefont{Marko}},
  \bibinfo{author}{\bibfnamefont{E.~D.} \bibnamefont{Siggia}},
  \bibnamefont{and} \bibinfo{author}{\bibfnamefont{S.}~\bibnamefont{Smith}},
  \bibinfo{journal}{Science} \textbf{\bibinfo{volume}{265}},
  \bibinfo{pages}{1599} (\bibinfo{year}{1994}).

\bibitem[{\citenamefont{Marko and Siggia}(1995)}]{Marko1995}
\bibinfo{author}{\bibfnamefont{J.~F.} \bibnamefont{Marko}} \bibnamefont{and}
  \bibinfo{author}{\bibfnamefont{E.~D.} \bibnamefont{Siggia}},
  \bibinfo{journal}{Macromolecules} \textbf{\bibinfo{volume}{28}},
  \bibinfo{pages}{8759} (\bibinfo{year}{1995}).

\bibitem[{\citenamefont{Shore and Baldwin}(1983)}]{Shore1983}
\bibinfo{author}{\bibfnamefont{D.}~\bibnamefont{Shore}} \bibnamefont{and}
  \bibinfo{author}{\bibfnamefont{R.~L.} \bibnamefont{Baldwin}},
  \bibinfo{journal}{J. Mol. Biol.} \textbf{\bibinfo{volume}{170}},
  \bibinfo{pages}{957} (\bibinfo{year}{1983}).

\bibitem[{\citenamefont{Olson et~al.}(1998)\citenamefont{Olson, Gorin, Lu,
  Hock, and Zhurkin}}]{Olson1998}
\bibinfo{author}{\bibfnamefont{W.~K.} \bibnamefont{Olson}},
  \bibinfo{author}{\bibfnamefont{A.~A.} \bibnamefont{Gorin}},
  \bibinfo{author}{\bibfnamefont{X.-J.} \bibnamefont{Lu}},
  \bibinfo{author}{\bibfnamefont{L.~M.} \bibnamefont{Hock}}, \bibnamefont{and}
  \bibinfo{author}{\bibfnamefont{V.~B.} \bibnamefont{Zhurkin}},
  \bibinfo{journal}{Proc. Natl. Acad. Sci. U. S. A.}
  \textbf{\bibinfo{volume}{95}}, \bibinfo{pages}{11163} (\bibinfo{year}{1998}).

\bibitem[{\citenamefont{Scipioni et~al.}(2002)\citenamefont{Scipioni, Anselmi,
  Zuccheri, Samori, and Santis}}]{Scipioni2002}
\bibinfo{author}{\bibfnamefont{A.}~\bibnamefont{Scipioni}},
  \bibinfo{author}{\bibfnamefont{C.}~\bibnamefont{Anselmi}},
  \bibinfo{author}{\bibfnamefont{G.}~\bibnamefont{Zuccheri}},
  \bibinfo{author}{\bibfnamefont{B.}~\bibnamefont{Samori}}, \bibnamefont{and}
  \bibinfo{author}{\bibfnamefont{P.~D.} \bibnamefont{Santis}},
  \bibinfo{journal}{Biophys. J.} \textbf{\bibinfo{volume}{83}},
  \bibinfo{pages}{2408} (\bibinfo{year}{2002}).

\bibitem[{\citenamefont{Garcia et~al.}(2007)\citenamefont{Garcia, Grayson, Han,
  Inamdar, Kondev, Nelson, Phillips, Widom, and Wiggins}}]{Garcia2007a}
\bibinfo{author}{\bibfnamefont{H.~G.} \bibnamefont{Garcia}},
  \bibinfo{author}{\bibfnamefont{P.}~\bibnamefont{Grayson}},
  \bibinfo{author}{\bibfnamefont{L.}~\bibnamefont{Han}},
  \bibinfo{author}{\bibfnamefont{M.}~\bibnamefont{Inamdar}},
  \bibinfo{author}{\bibfnamefont{J.}~\bibnamefont{Kondev}},
  \bibinfo{author}{\bibfnamefont{P.~C.} \bibnamefont{Nelson}},
  \bibinfo{author}{\bibfnamefont{R.}~\bibnamefont{Phillips}},
  \bibinfo{author}{\bibfnamefont{J.}~\bibnamefont{Widom}}, \bibnamefont{and}
  \bibinfo{author}{\bibfnamefont{P.~A.} \bibnamefont{Wiggins}},
  \bibinfo{journal}{Biopolymers} \textbf{\bibinfo{volume}{85}},
  \bibinfo{pages}{115} (\bibinfo{year}{2007}).

\bibitem[{\citenamefont{Geanacopoulos et~al.}(2001)\citenamefont{Geanacopoulos,
  Vasmatzis, Zhurkin, and Adhya}}]{Geanacopoulos2001}
\bibinfo{author}{\bibfnamefont{M.}~\bibnamefont{Geanacopoulos}},
  \bibinfo{author}{\bibfnamefont{G.}~\bibnamefont{Vasmatzis}},
  \bibinfo{author}{\bibfnamefont{V.~B.} \bibnamefont{Zhurkin}},
  \bibnamefont{and} \bibinfo{author}{\bibfnamefont{S.}~\bibnamefont{Adhya}},
  \bibinfo{journal}{Nature Struct. Biol.} \textbf{\bibinfo{volume}{8}},
  \bibinfo{pages}{432} (\bibinfo{year}{2001}).

\bibitem[{\citenamefont{Jaffar~Ali et~al.}(2001)\citenamefont{Jaffar~Ali, Amit,
  Braslavsky, Oppenheim, Gileadii, and Stavans}}]{Ali2001}
\bibinfo{author}{\bibfnamefont{B.~M.} \bibnamefont{Jaffar~Ali}},
  \bibinfo{author}{\bibfnamefont{R.}~\bibnamefont{Amit}},
  \bibinfo{author}{\bibfnamefont{I.}~\bibnamefont{Braslavsky}},
  \bibinfo{author}{\bibfnamefont{A.~B.} \bibnamefont{Oppenheim}},
  \bibinfo{author}{\bibfnamefont{O.}~\bibnamefont{Gileadii}}, \bibnamefont{and}
  \bibinfo{author}{\bibfnamefont{J.}~\bibnamefont{Stavans}},
  \bibinfo{journal}{Proc. Natl. Acad. Sci. U. S. A.}
  \textbf{\bibinfo{volume}{98}}, \bibinfo{pages}{10658} (\bibinfo{year}{2001}).

\bibitem[{\citenamefont{Amit et~al.}(2003)\citenamefont{Amit, Oppenheim, and
  Stavans}}]{Amit2003}
\bibinfo{author}{\bibfnamefont{R.}~\bibnamefont{Amit}},
  \bibinfo{author}{\bibfnamefont{A.~B.} \bibnamefont{Oppenheim}},
  \bibnamefont{and} \bibinfo{author}{\bibfnamefont{J.}~\bibnamefont{Stavans}},
  \bibinfo{journal}{Biophys. J.} \textbf{\bibinfo{volume}{84}},
  \bibinfo{pages}{2467} (\bibinfo{year}{2003}).

\bibitem[{\citenamefont{van Noort et~al.}(2004)\citenamefont{van Noort,
  Verbrugge, Goosen, Dekker, and Dame}}]{Noort2004}
\bibinfo{author}{\bibfnamefont{J.}~\bibnamefont{van Noort}},
  \bibinfo{author}{\bibfnamefont{S.}~\bibnamefont{Verbrugge}},
  \bibinfo{author}{\bibfnamefont{N.}~\bibnamefont{Goosen}},
  \bibinfo{author}{\bibfnamefont{C.}~\bibnamefont{Dekker}}, \bibnamefont{and}
  \bibinfo{author}{\bibfnamefont{R.~T.} \bibnamefont{Dame}},
  \bibinfo{journal}{Proc. Natl. Acad. Sci. U. S. A.}
  \textbf{\bibinfo{volume}{101}}, \bibinfo{pages}{6969} (\bibinfo{year}{2004}).

\bibitem[{\citenamefont{Du et~al.}(2005{\natexlab{a}})\citenamefont{Du,
  Vologodskaia, Kuhn, Frank-Kamenetskii, and Vologodskii}}]{Du2005a}
\bibinfo{author}{\bibfnamefont{Q.}~\bibnamefont{Du}},
  \bibinfo{author}{\bibfnamefont{M.}~\bibnamefont{Vologodskaia}},
  \bibinfo{author}{\bibfnamefont{H.}~\bibnamefont{Kuhn}},
  \bibinfo{author}{\bibfnamefont{M.}~\bibnamefont{Frank-Kamenetskii}},
  \bibnamefont{and}
  \bibinfo{author}{\bibfnamefont{A.}~\bibnamefont{Vologodskii}},
  \bibinfo{journal}{Biophys. J.} \textbf{\bibinfo{volume}{88}},
  \bibinfo{pages}{4137} (\bibinfo{year}{2005}{\natexlab{a}}).

\bibitem[{\citenamefont{Yuan et~al.}(2006)\citenamefont{Yuan, Rhoades, Lou, and
  Archer}}]{Archer2006}
\bibinfo{author}{\bibfnamefont{C.}~\bibnamefont{Yuan}},
  \bibinfo{author}{\bibfnamefont{E.}~\bibnamefont{Rhoades}},
  \bibinfo{author}{\bibfnamefont{X.~W.} \bibnamefont{Lou}}, \bibnamefont{and}
  \bibinfo{author}{\bibfnamefont{L.~A.} \bibnamefont{Archer}},
  \bibinfo{journal}{Nucleic Acids Res.} \textbf{\bibinfo{volume}{34}},
  \bibinfo{pages}{4554} (\bibinfo{year}{2006}).

\bibitem[{\citenamefont{Cloutier and Widom}(2004)}]{Cloutier2004}
\bibinfo{author}{\bibfnamefont{T.~E.} \bibnamefont{Cloutier}} \bibnamefont{and}
  \bibinfo{author}{\bibfnamefont{J.}~\bibnamefont{Widom}},
  \bibinfo{journal}{Mol. Cell} \textbf{\bibinfo{volume}{14}},
  \bibinfo{pages}{355} (\bibinfo{year}{2004}).

\bibitem[{\citenamefont{Wiggins et~al.}(2006)\citenamefont{Wiggins, van~der
  Heijden, Moreno-Herrero, Spakowitz, Phillips, Widom, Dekker, and
  Nelson}}]{Wiggins2006Nature}
\bibinfo{author}{\bibfnamefont{P.~A.} \bibnamefont{Wiggins}},
  \bibinfo{author}{\bibfnamefont{T.}~\bibnamefont{van~der Heijden}},
  \bibinfo{author}{\bibfnamefont{F.}~\bibnamefont{Moreno-Herrero}},
  \bibinfo{author}{\bibfnamefont{A.}~\bibnamefont{Spakowitz}},
  \bibinfo{author}{\bibfnamefont{R.}~\bibnamefont{Phillips}},
  \bibinfo{author}{\bibfnamefont{J.}~\bibnamefont{Widom}},
  \bibinfo{author}{\bibfnamefont{C.}~\bibnamefont{Dekker}}, \bibnamefont{and}
  \bibinfo{author}{\bibfnamefont{P.~C.} \bibnamefont{Nelson}},
  \bibinfo{journal}{Nature Nanotech.} \textbf{\bibinfo{volume}{1}},
  \bibinfo{pages}{137} (\bibinfo{year}{2006}).

\bibitem[{\citenamefont{Shroff et~al.}(2005)\citenamefont{Shroff, Reinhard,
  Siu, Agarwal, Spakowitz, and Liphardt}}]{Shroff2005}
\bibinfo{author}{\bibfnamefont{H.}~\bibnamefont{Shroff}},
  \bibinfo{author}{\bibfnamefont{B.}~\bibnamefont{Reinhard}},
  \bibinfo{author}{\bibfnamefont{M.}~\bibnamefont{Siu}},
  \bibinfo{author}{\bibfnamefont{H.}~\bibnamefont{Agarwal}},
  \bibinfo{author}{\bibfnamefont{A.}~\bibnamefont{Spakowitz}},
  \bibnamefont{and} \bibinfo{author}{\bibfnamefont{J.}~\bibnamefont{Liphardt}},
  \bibinfo{journal}{Nano Lett.} \textbf{\bibinfo{volume}{5}},
  \bibinfo{pages}{1509} (\bibinfo{year}{2005}).

\bibitem[{\citenamefont{Du et~al.}(2007)\citenamefont{Du, Kotlyar, and
  Vologodskii}}]{Du2007}
\bibinfo{author}{\bibfnamefont{Q.}~\bibnamefont{Du}},
  \bibinfo{author}{\bibfnamefont{A.}~\bibnamefont{Kotlyar}}, \bibnamefont{and}
  \bibinfo{author}{\bibfnamefont{A.}~\bibnamefont{Vologodskii}},
  \bibinfo{journal}{Nucleic Acids Res.}  (\bibinfo{year}{2007}).

\bibitem[{\citenamefont{Yan and Marko}(2004)}]{Yan2004}
\bibinfo{author}{\bibfnamefont{J.}~\bibnamefont{Yan}} \bibnamefont{and}
  \bibinfo{author}{\bibfnamefont{J.~F.} \bibnamefont{Marko}},
  \bibinfo{journal}{Phys. Rev. Lett.} \textbf{\bibinfo{volume}{93}},
  \bibinfo{pages}{108108} (\bibinfo{year}{2004}).

\bibitem[{\citenamefont{Wiggins et~al.}(2005)\citenamefont{Wiggins, Phillips,
  and Nelson}}]{Wiggins2005}
\bibinfo{author}{\bibfnamefont{P.~A.} \bibnamefont{Wiggins}},
  \bibinfo{author}{\bibfnamefont{R.}~\bibnamefont{Phillips}}, \bibnamefont{and}
  \bibinfo{author}{\bibfnamefont{P.~C.} \bibnamefont{Nelson}},
  \bibinfo{journal}{Phys. Rev. E} \textbf{\bibinfo{volume}{71}},
  \bibinfo{pages}{021909} (\bibinfo{year}{2005}).

\bibitem[{\citenamefont{Kahn et~al.}(1994)\citenamefont{Kahn, Yun, and
  Crothers}}]{Kahn1994}
\bibinfo{author}{\bibfnamefont{J.~D.} \bibnamefont{Kahn}},
  \bibinfo{author}{\bibfnamefont{E.}~\bibnamefont{Yun}}, \bibnamefont{and}
  \bibinfo{author}{\bibfnamefont{D.~M.} \bibnamefont{Crothers}},
  \bibinfo{journal}{Nature} \textbf{\bibinfo{volume}{368}},
  \bibinfo{pages}{163} (\bibinfo{year}{1994}).

\bibitem[{\citenamefont{Kahn and Crothers}(1992)}]{Kahn1992}
\bibinfo{author}{\bibfnamefont{J.~D.} \bibnamefont{Kahn}} \bibnamefont{and}
  \bibinfo{author}{\bibfnamefont{D.~M.} \bibnamefont{Crothers}},
  \bibinfo{journal}{Proc. Natl. Acad. Sci. U. S. A.}
  \textbf{\bibinfo{volume}{89}}, \bibinfo{pages}{6343} (\bibinfo{year}{1992}).

\bibitem[{\citenamefont{Yan and Marko}(2003)}]{Yan2003}
\bibinfo{author}{\bibfnamefont{J.}~\bibnamefont{Yan}} \bibnamefont{and}
  \bibinfo{author}{\bibfnamefont{J.~F.} \bibnamefont{Marko}},
  \bibinfo{journal}{Phys. Rev. E} \textbf{\bibinfo{volume}{68}},
  \bibinfo{pages}{011905} (\bibinfo{year}{2003}).

\bibitem[{\citenamefont{Soennichsen et~al.}(2005)\citenamefont{Soennichsen,
  Reinhard, Liphardt, and Alivisatos}}]{Sonnichsen2005}
\bibinfo{author}{\bibfnamefont{C.}~\bibnamefont{Soennichsen}},
  \bibinfo{author}{\bibfnamefont{B.}~\bibnamefont{Reinhard}},
  \bibinfo{author}{\bibfnamefont{J.}~\bibnamefont{Liphardt}}, \bibnamefont{and}
  \bibinfo{author}{\bibfnamefont{A.}~\bibnamefont{Alivisatos}},
  \bibinfo{journal}{Nature Biotech.} \textbf{\bibinfo{volume}{23}},
  \bibinfo{pages}{741} (\bibinfo{year}{2005}).

\bibitem[{\citenamefont{Ranjith et~al.}(2005)\citenamefont{Ranjith, Kumar, and
  Menon}}]{Ranjith2005}
\bibinfo{author}{\bibfnamefont{P.}~\bibnamefont{Ranjith}},
  \bibinfo{author}{\bibfnamefont{P.~B.~S.} \bibnamefont{Kumar}},
  \bibnamefont{and} \bibinfo{author}{\bibfnamefont{G.}~\bibnamefont{Menon}},
  \bibinfo{journal}{Phys. Rev. Lett.} \textbf{\bibinfo{volume}{94}},
  \bibinfo{pages}{138102} (\bibinfo{year}{2005}).

\bibitem[{\citenamefont{Yan et~al.}(2005)\citenamefont{Yan, Kawamura, and
  Marko}}]{Yan2005}
\bibinfo{author}{\bibfnamefont{J.}~\bibnamefont{Yan}},
  \bibinfo{author}{\bibfnamefont{R.}~\bibnamefont{Kawamura}}, \bibnamefont{and}
  \bibinfo{author}{\bibfnamefont{J.~F.} \bibnamefont{Marko}},
  \bibinfo{journal}{Phys. Rev. E} \textbf{\bibinfo{volume}{71}},
  \bibinfo{pages}{061905} (\bibinfo{year}{2005}).

\bibitem[{\citenamefont{Douarche and Cocco}(2005)}]{Douarche2005}
\bibinfo{author}{\bibfnamefont{N.}~\bibnamefont{Douarche}} \bibnamefont{and}
  \bibinfo{author}{\bibfnamefont{S.}~\bibnamefont{Cocco}},
  \bibinfo{journal}{Physical Review E} \textbf{\bibinfo{volume}{72}},
  \bibinfo{pages}{061902} (\bibinfo{year}{2005}).

\bibitem[{\citenamefont{Seol et~al.}(2007)\citenamefont{Seol, Li, Nelson,
  Perkins, and Betterton}}]{Seol2007}
\bibinfo{author}{\bibfnamefont{Y.}~\bibnamefont{Seol}},
  \bibinfo{author}{\bibfnamefont{J.}~\bibnamefont{Li}},
  \bibinfo{author}{\bibfnamefont{P.~C.} \bibnamefont{Nelson}},
  \bibinfo{author}{\bibfnamefont{T.~T.} \bibnamefont{Perkins}},
  \bibnamefont{and} \bibinfo{author}{\bibfnamefont{M.~D.}
  \bibnamefont{Betterton}}, \bibinfo{journal}{Biophys. J.}
  \textbf{\bibinfo{volume}{93}}, \bibinfo{pages}{4360} (\bibinfo{year}{2007}).

\bibitem[{\citenamefont{Neumann}(2003)}]{Neumann2003}
\bibinfo{author}{\bibfnamefont{R.~M.} \bibnamefont{Neumann}},
  \bibinfo{journal}{Biophys. J.} \textbf{\bibinfo{volume}{85}},
  \bibinfo{pages}{3418} (\bibinfo{year}{2003}).

\bibitem[{\citenamefont{Wilhelm and Frey}(1996)}]{Wilhelm1996}
\bibinfo{author}{\bibfnamefont{J.}~\bibnamefont{Wilhelm}} \bibnamefont{and}
  \bibinfo{author}{\bibfnamefont{E.}~\bibnamefont{Frey}},
  \bibinfo{journal}{Phys. Rev. Lett.} \textbf{\bibinfo{volume}{77}},
  \bibinfo{pages}{2581} (\bibinfo{year}{1996}).

\bibitem[{\citenamefont{Shore et~al.}(1981)\citenamefont{Shore, Langowski, and
  Baldwin}}]{Shore1981}
\bibinfo{author}{\bibfnamefont{D.}~\bibnamefont{Shore}},
  \bibinfo{author}{\bibfnamefont{J.}~\bibnamefont{Langowski}},
  \bibnamefont{and} \bibinfo{author}{\bibfnamefont{R.~L.}
  \bibnamefont{Baldwin}}, \bibinfo{journal}{Proc. Natl. Acad. Sci. U. S. A.}
  \textbf{\bibinfo{volume}{78}}, \bibinfo{pages}{4833} (\bibinfo{year}{1981}).

\bibitem[{\citenamefont{Shimada and Yamakawa}(1984)}]{Yamakawa1984}
\bibinfo{author}{\bibfnamefont{J.}~\bibnamefont{Shimada}} \bibnamefont{and}
  \bibinfo{author}{\bibfnamefont{H.}~\bibnamefont{Yamakawa}},
  \bibinfo{journal}{Macromolecules} \textbf{\bibinfo{volume}{17}},
  \bibinfo{pages}{689} (\bibinfo{year}{1984}).

\bibitem[{\citenamefont{Podtelezhnikov and
  Vologodskii}(2000)}]{Vologodskii2000}
\bibinfo{author}{\bibfnamefont{A.}~\bibnamefont{Podtelezhnikov}}
  \bibnamefont{and}
  \bibinfo{author}{\bibfnamefont{A.}~\bibnamefont{Vologodskii}},
  \bibinfo{journal}{Macromolecules} \textbf{\bibinfo{volume}{33}},
  \bibinfo{pages}{2767} (\bibinfo{year}{2000}).

\bibitem[{\citenamefont{Zhang and Crothers}(2003)}]{Zhang2003}
\bibinfo{author}{\bibfnamefont{Y.}~\bibnamefont{Zhang}} \bibnamefont{and}
  \bibinfo{author}{\bibfnamefont{D.~M.} \bibnamefont{Crothers}},
  \bibinfo{journal}{Biophys. J.} \textbf{\bibinfo{volume}{84}},
  \bibinfo{pages}{136} (\bibinfo{year}{2003}).

\bibitem[{\citenamefont{Sankararaman and Marko}(2005)}]{Sankararaman2005}
\bibinfo{author}{\bibfnamefont{S.}~\bibnamefont{Sankararaman}}
  \bibnamefont{and} \bibinfo{author}{\bibfnamefont{J.~F.} \bibnamefont{Marko}},
  \bibinfo{journal}{Phys. Rev. E} \textbf{\bibinfo{volume}{71}},
  \bibinfo{pages}{021911} (\bibinfo{year}{2005}).

\bibitem[{\citenamefont{Tkachenko}(2007)}]{Tkachenko2007}
\bibinfo{author}{\bibfnamefont{A.~V.} \bibnamefont{Tkachenko}},
  \bibinfo{journal}{arXiv:q-bio/0703026v1}  (\bibinfo{year}{2007}).

\bibitem[{\citenamefont{Du et~al.}(2005{\natexlab{b}})\citenamefont{Du, Smith,
  Shiffeldrim, Vologodskaia, and Vologodskii}}]{Du2005}
\bibinfo{author}{\bibfnamefont{Q.}~\bibnamefont{Du}},
  \bibinfo{author}{\bibfnamefont{C.}~\bibnamefont{Smith}},
  \bibinfo{author}{\bibfnamefont{N.}~\bibnamefont{Shiffeldrim}},
  \bibinfo{author}{\bibfnamefont{M.}~\bibnamefont{Vologodskaia}},
  \bibnamefont{and}
  \bibinfo{author}{\bibfnamefont{A.}~\bibnamefont{Vologodskii}},
  \bibinfo{journal}{Proc. Natl. Acad. Sci. U. S. A.}
  \textbf{\bibinfo{volume}{102}}, \bibinfo{pages}{5397}
  (\bibinfo{year}{2005}{\natexlab{b}}).

\end{thebibliography}

\end{document}